# On discriminating between Libby-Novick generalized beta and Kumaraswamy distributions: theory and methods

Indranil Ghosh

# On discriminating between Libby-Novick generalized beta and Kumaraswamy distributions: theory and methods

Indranil Ghosh

Department of Mathematics and Statistics, University of North Carolina, Wilmington, Wilmington, NC, USA

**ABSTRACT**

In fitting a continuous bounded data, the generalized beta (and several variants of this distribution) and the two-parameter Kumaraswamy (KW) distributions are the two most prominent univariate continuous distributions that come to our mind. There are some common features between these two rival probability models and to select one of them in a practical situation can be of great interest. Consequently, in this paper, we discuss various methods of selection between the generalized beta proposed by Libby and Novick (1982) (LNGB) and the KW distributions, such as the criteria based on probability of correct selection which is an improvement over the likelihood ratio statistic approach, and also based on pseudo-distance measures. We obtain an approximation for the probability of correct selection under the hypotheses $H_{LNGB}$ and $H_{KW}$, and select the model that maximizes it. However, our proposal is more appealing in the sense that we provide the comparison study for the LNGB distribution that subsumes both types of classical beta and exponentiated generators (see, for details, Cordeiro et al. 2014; Libby and Novick 1982) which can be a natural competitor of a two-parameter KW distribution in an appropriate scenario.

## 1. Introduction

Methods of discriminating between two or more probability models (in both continuous and in the discrete domains) are not new in the literature, and it has several useful applications, including, but not limited to, design of experiments, see Atkinson (1970) and the references cited therein. Atkinson and Fedorov (1975) discussed a strategy in which a combined distribution containing the component models as special cases are investigated by constructing summary statistics based on the combined distribution and then formulate a test of departures from one model to the other. Balakrishnan and Ristić (2016) discussed the role of maximum entropy for the selection of parent distribution among two parent distributions, as a procedure for discrimination between two probability models. In this paper, we consider the problem of discriminating between two bounded (defined on (0, 1)) absolutely continuous probability models, namely a two-parameter Kumaraswamy distribution (see, for details, Kumaraswamy (1980)), and a three parameter Libby-Novick generalized beta distribution which subsumes both types of classical beta distribution. However, Kumaraswamy argued that the beta distribution and its several generalizations does not faithfully fit hydrological random variable such as daily rainfall and daily stream flow. This motivates our current work in which we want to discriminate between a LNGB and a KW distribution by using the methodology of computing probability of correct selection. In addition, we adopt a new strategy for this purpose which appears to be more efficient that is based on several pseudo-distance measures.

It is interesting to note that under certain parametric conditions, both of these two probability models (i.e., KW and LNGB) reduces to a standard uniform distribution, while the LNGB distribution reduces to a Beta (type-I) distribution if one of the shape parameters takes the value 1. Therefore, in a practical setting, where the data points such as proportions, it will be interesting to see when the data indicates that the estimated values of the parameters closely resembles to the specific conditions for which both of them reduces to either a uniform or beta; and in such a scenario, how efficiently one can distinguish between them. Next, we provide the mathematical expression for the two probability models understudy:

- The probability density function (pdf) of a KW distribution with two shape parameters $\alpha > 0$ and $\delta > 0$ is defined by

$$f_{KW}(x) = \alpha \delta x^{\alpha-1}(1-x^\alpha)^{\delta-1}; \quad \text{and} \qquad (1)$$

$$F(x) = 1 - (1-x^\alpha)^\delta I(0 < x < 1). \qquad (2)$$

- The pdf of a three-parameter LNGB distribution is given by

$$f_{LNGB}(x) = \frac{\beta^a}{B(a,b)} x^{a-1}(1-x)^{b-1}[1-(1-\beta)x]^{-(a+b)} \qquad (3)$$

$$I(0 < x < 1), \qquad (4)$$

where $(a, b, \beta) \in \mathbb{R}^+$, and $\beta$ is the shape parameter, and $(a, b)$ are two scale parameters.

CONTACT Indranil Ghosh ghoshi@uncw.edu Department of Mathematics and Statistics, University of North Carolina, Wilmington, Wilmington, NC, USA.

Our main objective paper is to propose and discuss a probability based selection criterion and also a pseudo-divergence measure based criterion that has not been discussed before in the literature to the best of our knowledge which will be utilized to distinguish/discriminate between the two probability models as given in Eqs. (1) and (3) within the framework of two non-nested statistical hypothesis. The proposed criterion is based on the asymptotic distribution of the likelihood ratio statistic proposed by Cox (1961, 1962). With this, we obtain the probability of correct selection under the null hypotheses that the data comes from either the two-parameter KW or a three-parameter LNGB distribution and select the probability model that maximizes this probability of correct selection. The test statistic is the logarithm of the ratio of the maximized log-likelihoods under both the null and alternative hypotheses. This statistic is compared with its expected value under the null hypothesis. Small deviations of the expected mean imply evidences in favor of the null hypothesis, while large deviations indicate evidences against. Regularity conditions and a rigorous proof of the asymptotic normality of the Cox's test statistic was provided by White (1982a). In the literature, several authors have worked on this topic, a non-exhaustive list of such works can be found in, for instance, in the works of Bain and Engelhardt (1980), Fearn and Nebenzahl (1991), Gupta and Kundu (2004), Kundu, Gupta, and Manglick (2005), Dey and Kundu (2012), Ristić et al. (2018) and Silva et al. (2014). Here, we provide several pseudo-distance measures and minimum sample size criterion which is the new contribution in this topic. The remainder of the paper is organized as follows. In Section 2, we discuss the strategy of discriminating between the two probability models given in Eqs. (1) and (3) based on the log-likelihood ratio statistic and its' asymptotic null distribution approach. In Section 3, we discuss the probability of selection criterion based on the material developed in Section 2. Section 4 deals with the discussion a new strategy that is based on several pseudo-distance measures and associated minimum sample size criterion. In Section 5, we provide a lay out for undertaking a simulation study, and a small simulation study is presented to illustrate the feasibility of the proposed methodology. In Section 6, two real-life data sets are re-analyzed to illustrate the efficacy of the proposed methodology of probability of correct selection. Finally, some concluding remarks are made in Section 7.

## 2. Discrimination between the LNGB and the Kumaraswamy probability models

Let $\{X_i\}_{i=1}^n$ be i.i.d. with observed values $X_1, \ldots, X_n$ from a $LNGB(a, b, \beta)$ or $KW(\alpha, \delta)$ distribution with respective densities. We define

$$H_{LNGB} : \{X_i\}_{i=1}^n \sim LNGB(a, b, \beta)$$
$$H_{KW} : \{X_i\}_{i=1}^n \sim KW(\alpha, \delta).$$

Then, the log-likelihood function associated to the LNGB distribution is given by

$$\ell_{LNGB}(a, b, \beta) = na \log \beta - n \log B(a, b) + (a - 1) \sum_{i=1}^n \log X_i$$

$$+ (b - 1) \sum_{i=1}^n \log (1 - X_i)$$

$$- (a + b) \sum_{i=1}^n \log (1 - (1 - \beta)X_i). \quad (5)$$

From Eq. (5), the MLEs $(\widehat{a}_n, \widehat{b}_n, \widehat{\beta}_n)$ of $(a, b, \beta)$, respectively are obtained as solutions of the non-linear equations that are given as follows:

$$\frac{\partial \ell_{LNGB}(a, b, \beta)}{\partial a} = n \log \widehat{\beta}_n - n \left[ \Psi(\widehat{a}_n) - \Psi(\widehat{a}_n + \widehat{b}_n) \right]$$
$$+ \sum_{i=1}^n \log X_i - \sum_{i=1}^n \log \left(1 - (1 - \widehat{\beta}_n)X_i\right) = 0. \quad (6)$$

$$\frac{\partial \ell_{LNGB}(a, b, \beta)}{\partial b} = n \left[ \Psi(\widehat{b}_n) - \Psi(\widehat{a}_n + \widehat{b}_n) \right] - \sum_{i=1}^n \log(1 - X_i)$$
$$+ \sum_{i=1}^n \log \left(1 - (1 - \widehat{\beta}_n)X_i\right) = 0, \quad (7)$$

where $\Psi^{(m)}(x) = \frac{d^m \log \Gamma(x)}{d^{m+1}}$.

$$\frac{\partial \ell_{LNGB}(a, b, \beta)}{\partial \beta} \quad (8)$$
$$= n \frac{\widehat{a}_n}{\widehat{\beta}_n} - (\widehat{a}_n + \widehat{b}_n) \sum_{i=1}^n X_i \left[ \left(1 - (1 - \widehat{\beta}_n) X_i\right) \right]^{-1} = 0. \quad (9)$$

On the other hand, the log-likelihood function associated with the KW distribution is given by

$$\ell_{KW}(\alpha, \delta) = n \log \alpha + n \log \delta + (\alpha - 1) \sum_{i=1}^n \log X_i \quad (10)$$
$$+ (\delta - 1) \sum_{i=1}^n \log \left[1 - X_i^\alpha\right].$$

Therefore, the associated MLEs $(\widehat{\alpha}_n, \widehat{\delta}_n)$ of $(\alpha, \delta)$ are given by

$$\frac{\partial \ell_{KW}(\alpha, \delta)}{\partial \alpha} = \sum_{i=1}^n \log [X_i] + \frac{n}{\widehat{\alpha}_n} - (\widehat{\delta}_n - 1) \sum_{i=1}^n \frac{X_i^{\widehat{\alpha}_n} \log X_i}{1 - X_i^{\widehat{\alpha}_n}} = 0. \quad (11)$$

$$\frac{\partial \ell_{KW}(\alpha, \delta)}{\partial \delta} = \frac{n}{\widehat{\delta}_n} + \sum_{i=1}^n \log \left(1 - X_i^{\widehat{\alpha}_n}\right) = 0. \quad (12)$$

Next, we define our test statistic as

$$W_n = \log \left[ \frac{\prod_{i=1}^n f_{LNGB}\left(X_i, \widehat{a}_n, \widehat{b}_n, \widehat{\beta}_n\right)}{\prod_{i=1}^n f_{KW}\left(X_i, \widehat{\alpha}_n, \widehat{\delta}_n\right)} \right] \quad (13)$$
$$= \ell_{LNGB}(a, b, \beta) - \ell_{KW}(\alpha, \delta).$$

More explicitly, our test statistic can be written as

$$W_n = n \left[ -\log B(\widehat{a}_n, \widehat{b}_n) - \log (\widehat{\alpha}_n \widehat{\delta}_n) + \widehat{a}_n \log \widehat{\beta}_n \right]$$

$$-n\left[1 - \widehat{\beta}_n^{-1}\right] + \sum_{i=1}^{n} \log(1 - X_i) \left[\widehat{a}_n + 2\widehat{b}_n - 1\right]$$

$$+ \sum_{i=1}^{n} \log(X_i) \left[\widehat{a}_n - \widehat{\alpha}_n\right]$$

$$- \left(\widehat{a}_n + \widehat{b}_n\right) \left[-\Psi\left(\widehat{a}_n + \widehat{b}_n\right) + \Psi(\widehat{b}_n)\right], \quad (14)$$

on using Eqs. (6)–(8) and Eqs. (11)–(12), respectively. Next, we consider the two cases separately, in each case we consider LNGB (KW) as the true population distribution while the other one as the distribution under the alternative hypothesis. In practical situations, one may consider the Akaike Information Criterion (AIC), or the Bayesian Information Criterion (BIC) as a selection criteria. However, in this paper, we here consider a different selection criterion that is based on the asymptotic distribution of a normalized version of the test statistics $W_n$ under the hypotheses $H_{LNGB}$ and $H_{KW}$. This details will be discussed in the next two subsections. Next, we begin our discussion by focusing on deriving the asymptotic null distribution of the test statistic in two specific scenarios: (i) when the true data distribution is a three parameter LNGB; and (ii) when the true data distribution is a two-parameter KW.

### 2.1. Situation when the LNGB distribution is the null hypothesis

Here our goal is to find the asymptotic distribution of $W_n$ under the null hypothesis $H_{LNGB}$ against the alternative $H_{KW}$. We provide some useful mathematical preliminaries along this process which are given as follows. Let us suppose that $X_1, \ldots, X_n \sim LNGB(a, b, \beta)$. For any Borel measurable function $h()$, the under subscript LNGB in $E_{\mathbb{LNGB}}(h(X))$ will represent the following:

$$E_{\mathbb{LNGB}}(h(X)) = \int_0^1 h(x) f_{LNGB}(x; a, b, \beta) \, dx.$$

Observe that under the null hypothesis $H_{LNGB}$, as $n \to \infty$,

(i) $\widehat{a}_n \to a$, $\widehat{b}_n \to b$, and $\widehat{\beta}_n \to \beta$ almost surely where

$$E_{LNGB}\left[\log f_{LNGB}(X; a, b, \beta)\right]$$
$$= \max_{\bar{a}, \bar{b}, \bar{\beta}} E_{LNGB}\left[\log f_{LNGB}(X; \bar{a}, \bar{b}, \bar{\beta})\right].$$

(ii) $\widehat{\alpha}_n \to \tilde{\alpha}$, and $\widehat{\delta}_n \to \tilde{\delta}$ almost surely where

$$E_{LNGB}\left[\log f_{KW}(X; \tilde{\alpha}, \tilde{\delta})\right] = \max_{\alpha, \delta} E_{LNGB}\left[\log f_{KW}(X; \alpha, \delta)\right].$$

The maximum likelihood estimators $\tilde{\alpha}, \tilde{\delta}$ are functions of $a$ and $b$, and $\beta$ which is not further simplified in order to simplify the notation. The above convergence(s) follow from the results discussed in details by White (1982b). Next, in order to present the asymptotic distribution of the test statistic $W_n$ under $H_{LNGB}$, we need to compute the mean and variance of the random variable $W_n = \log f_{LNGB}(X; a, b, \beta) - \log f_{KW}(X; \tilde{\alpha}, \tilde{\delta})$ under the condition that $X \sim LNGB(a, b, \beta)$ which will be denoted by $M_{LNGB}(a, b, \beta)$ and $Var_{LNGB}(a, b, \beta)$, respectively, for which the details of their derivations are given in Appendix A.

**Table 1.** Values of $M_{LNGB}(a, b, \beta)$ and $Var_{LNGB}(a, b, \beta)$, $\tilde{\alpha}$ and $\tilde{\delta}$ for $a = 1.2, b = 1.5$, and some representative values of $\beta$.

| $\beta$ | $M_{LNGB}(a, b, \beta)$ | $Var_{LNGB}(a, b, \beta)$ | $\tilde{\alpha}$ | $\tilde{\delta}$ |
|---|---|---|---|---|
| 0.2 | −0.011825 | 0.746237 | 0.1626 | 3.0761 |
| 0.5 | −0.001315 | 0.071849 | 0.4549 | 2.2410 |
| 0.7 | −0.000259 | 0.014987 | 0.6667 | 2.0968 |
| 1.2 | −0.000037 | 0.002621 | 1.2292 | 1.9668 |
| 1.5 | −0.000143 | 0.010834 | 1.5801 | 1.9372 |
| 2.0 | −0.000294 | 0.025130 | 2.1773 | 1.9122 |

Next, we provide the following theorem which represents the asymptotic null distribution of $W_n$ under the null hypothesis $H_{LNGB}$.

*Theorem 1.* The asymptotic distribution of (under the null hypothesis $H_{LNGB}$), of $W_n$ is given by

$$n^{-1/2}(W_n - E_{LNGB}(W_n)) \sim n^{-1/2}\left(\tilde{W}_n^{LNGB} - nM_{LNGB}(a, b, \beta)\right)$$
$$\xrightarrow{D} N(0, Var_{LNGB}(a, b, \beta)), \quad (15)$$

as $n \to \infty$, where $\tilde{W}_n^{LNGB} = \ell_{KW}\left(\tilde{\alpha}, \tilde{\delta}\right) - \ell_{LNGB}(a, b, \beta)$.

*Proof.* From the Central Limit Theorem, it follows that $n^{-1/2}\left(\tilde{W}_n^{LNGB} - nM_{LNGB}(a, b, \beta)\right) \xrightarrow{D} N(0, Var_{LNGB}(a, b, \beta))$, as $n \to \infty$. Consequently, the remainder of the proof lies in showing the asymptotic equivalence between $n^{-1/2}(W_n - E_{LNGB}(W_{1n}))$ and $n^{-1/2}\left(\tilde{W}_n^{LNGB} - nM_{LNGB}(a, b, \beta)\right)$. This follows from an adaptation of the results presented in White (1982a). This completes the proof. □

In Table 1, some representative values of the $M_{LNGB}(a, b, \beta)$ and $Var_{LNGB}(a, b, \beta)$ for specific choices of $a$ and $b$ and for some choices of $\beta$ are provided. These values are for illustrative purposes as to how the mean and variance of the random variable $W_n$ varies for different (given) choices of the parameters $a, b, \beta$, and for estimated values of $\alpha$ and $\delta$ based on the procedure described earlier.

### 2.2. Scenario when the Kumaraswamy distribution is the null hypothesis

Here, we suppose that $X_1, \cdots, X_n \sim KW(\alpha, \delta)$, i.e., the true probability model for the data. As before, we consider some pertinent preliminaries given below.

Note that under the hypothesis $H_{KW}$, as $n \to \infty$,

(i) $\widehat{\alpha}_n \to \alpha$, and $\widehat{\delta}_n \to \delta$ almost surely, where

$$E_{KW}\left[\log f_{KW}(X; \alpha, \delta)\right] = \max_{\tilde{\alpha}, \tilde{\delta}} E_{KW}\left[\log f_{KW}(X; \tilde{\alpha}, \tilde{\delta})\right].$$

(ii) $\widehat{a}_n \to a$, and $\widehat{b}_n \to b$, and $\widehat{\beta}_n \to \beta$, almost surely, where

$$E_{KW}\left[\log f_{LNGB}(X; \widehat{a}, \widehat{b}, \widehat{\beta})\right]$$
$$= \max_{a, b, \beta} E_{LNGB}\left[\log f_{LNGB}\left(X; \widehat{a}, \widehat{b}, \widehat{\beta}\right)\right].$$

**Table 2.** Values of $M_{KW}(\alpha,\delta)$ and $Var_{KW}(\alpha,\delta)$, $\tilde{a}, \tilde{b}$, and $\tilde{\beta}$, and for $\delta = 2.4$ and some representative values of $\alpha$.

| $\alpha$ | $M_{KW}(\alpha,\delta)$ | $Var_{KW}(\alpha,\delta)$ | $\tilde{a}$ | $\tilde{b}$ | $\tilde{\beta}$ |
|---|---|---|---|---|---|
| 0.2 | −0.013467 | 0.08541 | 0.13283 | 1.2745 | 0.5678 |
| 0.5 | −0.02618 | 0.05231 | 0.5133 | 1.6853 | 1.4326 |
| 0.7 | −0.003619 | 0.03569 | 0.7323 | 1.1754 | 0.9846 |
| 1.2 | −0.004879 | 0.06783 | 1.0543 | 1.3234 | 2.3315 |
| 1.5 | −0.03947 | 0.08156 | 2.0128 | 2.0129 | 1.6748 |
| 2.0 | −0.06982 | 0.1948 | 1.0685 | 2.0732 | 1.0917 |

The quasi-maximum likelihood estimators $\widehat{a}, \widehat{b}, \widehat{\beta}$ are functions of $a$ and $b$, and $\beta$ which is not further simplified in order to simplify the notation. The above convergence(s) follow from the results discussed in details by White (1982b). Next, in order to present the asymptotic distribution of the test statistic $W_n$ under $H_{KW}$, we need to compute the mean and variance of the random variable $W_n = \log f_{LNGB}(X; \tilde{a}, \tilde{b}, \tilde{\beta}) - \log f_{KW}(X; \alpha, \delta)$ under the condition that $X \sim KW(X; \alpha, \delta)$ which will be denoted by $M_{KW}(\alpha, \delta)$ and $Var_{KW}(\alpha, \delta)$ respectively. For the details on their variation, see Appendix B.

Next, we provide the following theorem which represents the asymptotic null distribution of $W_n$ under the null hypothesis $H_{KW}$.

*Theorem 2.* The asymptotic distribution of (under the null hypothesis $H_{KW}$), of $W_n$ is given by

$$n^{-1/2}(W_n - E_{KW}(W_n)) \quad (16)$$
$$\sim n^{-1/2}\left(\tilde{W}_n^{KW} - nM_{KW}(\alpha,\delta)\right) \xrightarrow{D} N(0, Var_{KW}(\alpha,\delta)),$$

as $n \to \infty$, where $\tilde{W}_n^{KW} = \ell_{LNGB}(\tilde{a}, \tilde{b}, \tilde{\beta}) - \ell_{KW}((\alpha, \delta))$.

*Proof.* The proof is very similar to Theorem 1 and is omitted for brevity. □

In Table 2, some representative values of $M_{KW}(\alpha, \delta)$ and $Var_{KW}(\alpha, \delta)$ are listed for fixed value of $\delta$ for some representative values of $\alpha$. As before, these values are for illustrative purposes as to how the mean and variance of the random variable $W_n$ varies for different (given) choices of the parameters $\alpha$ and $\delta$ and for estimated values of $a, b, \beta$ based on the procedure described earlier.

Next, in the two sections, we discuss the two distinct strategies, viz., the probability based and via the pseudo-divergence and minimum sample size based as selection criteria in the context of choosing one of the two distributions assumed in this paper.

## 3. Probability based selection criterion

Let us first present asymptotic forms for the probabilities of correct selection (PCS) which are given for the two probability models as follows $PCS_{LNGB}(a, b, \beta) \equiv P(W_n < 0)$, and $PCS_{KW}(\alpha, \delta) \equiv P(W_n > 0)$, under the null hypotheses $H_{LNGB}$ and $H_{KW}$, respectively. Next, consider the scenario when the null and the alternative hypotheses are $H_{LNGB}$ and $H_{KW}$, respectively.

From the previous section, $PCS_{LNGB}(a, b, \beta)$ can be approximated by

$$PCS_{LNGB}(a,b,\beta) \equiv \Phi\left(\frac{\sqrt{n}M_{LNGB}(a,b,\beta)}{\sqrt{Var_{LNGB}(a,b,\beta)}}\right), \quad (17)$$

where $\Phi()$ is the cumulative distribution function of the standard normal distribution. Next, consider that the null and alternative hypotheses are $H_{KW}$ and $H_{LNGB}$ respectively. Subsequently, based on the convergence in distribution given in Eq. (16), $PCS_{KW}(\alpha, \delta)$ can be approximated by

$$PCS_{KW}(\alpha,\delta) \equiv \Phi\left(-\frac{\sqrt{n}M_{KW}(\alpha,\delta)}{\sqrt{Var_{KW}(\alpha,\delta)}}\right), \quad (18)$$

where the expressions for $M_{KW}(\alpha, \delta)$ and $Var_{KW}(a, b, \beta)$ are given in Appendix B.

Since the PCS as given in Eqs. (17) and (18) depends on the unknown parameters, in practice, we replace the parameters with their maximum likelihood estimators. Consequently, we define our selection criterion as follows:

(i) If $PCS_{LNGB}(\widehat{a}, \widehat{b}, \widehat{\beta}) < PCS_{KW}(\widehat{\alpha}, \widehat{\delta})$, choose the KW distribution, otherwise select the LNGB distribution, where the quantities under $\widehat{\ }$ are the maximum likelihood estimators of the respective parameters.

(ii) Equivalently, one can say that if

$$-M_{KW}(\alpha,\delta)\sqrt{Var_{LNGB}(a,b,\beta)}$$
$$> M_{LNGB}(a,b,\beta)\sqrt{Var_{KW}(\alpha,\delta)},$$

then select the KW distribution, otherwise select the LNGB distribution.

## 4. Distances and minimum sample size criterion

In this section, we discuss a procedure to determine the minimum sample size required in order to discriminate between the LNGB, and the KW distributions for a specific value(s) of the PCS and a given tolerance level that can be defined in terms of some pseudo-distance measures to determine the closeness between the two distributions understudy. There are several distance (or pseudo-distance) measures that are available in the literature to study the proximity between two probability distributions, among them, the Kolmogorov-Smirnov (KS), Hellinger distance are worthwhile to mention. For a detailed study on the use of pseudo-distance measures, see Cressie and Read (1984) and the references cited therein. Next, we provide some useful preliminaries in brief in this context.

Let $f$ and $g$ be two absolutely continuous density functions with the same support $\Omega = (0, 1)$ with distribution functions $F$ and $G$, respectively. Then,

- the Hellinger distance is given by

$$\mathbb{H}(f,g) = \int_\Omega \left[\sqrt{f(x)} - \sqrt{g(x)}\right]^2 dx = 2 - \int_\Omega \sqrt{f(x)g(x)}dx.$$

Next, we provide the following Lemmas which describes the expression of the Hellinger and the Power divergence distance for the two distributions as given in Eqs. (1) and (3).

*Lemma 1.* The Hellinger distance between the LNGB and the KW distribution will be

$$= 1 - \sqrt{\alpha\delta}\Gamma\left(\frac{a+\alpha}{2}\right)\beta^{\frac{a+b}{2}}\sqrt{\beta^{-a-b}}\Gamma\left(\frac{b+\delta}{2}\right)\sqrt{\frac{\beta^a}{B(a,b)}}$$
$$\times {}_2\tilde{F}_1\left(\frac{a+b}{2}, \frac{a+\alpha}{2}; \frac{1}{2}(a+b+\alpha+\delta); 1-\beta\right), \quad (19)$$

on using Mathematica.

- the Power divergence measure given by

$$PWD(f,g;\lambda) = \frac{1}{\lambda(\lambda+1)}\int_0^1 \left[\left[\frac{f(x)}{g(x)}\right]^\lambda - 1\right]f(x)dx. \quad (20)$$

Observe that for different choices of $\lambda$, it leads to several well-known divergence measures. For example, for $\lambda = -2, -1, -0.5, 0, 1$ implies Neyman Chi-square, Kullback-Leibler, squared Hellinger distance, Likelihood disparity, and Pearson Chi-square divergence respectively.

*Lemma 2.* The expression for the Power-divergence statistic between the LNGB and the KW distribution will be

$PWD(KW; LNGB; \lambda)$

$$= \frac{1}{\lambda(\lambda+1)}\Bigg[\alpha\delta\left((a\lambda - \alpha(\lambda+1))\Gamma(\lambda\alpha + \alpha + \delta - a\lambda - b\lambda + \delta\lambda + 1)\right)^{-1}$$

$$\times \Bigg[-\Gamma(\lambda\alpha + \alpha - a\lambda + 1)\beta^{-(\lambda(a+b))}\Gamma(\lambda\delta + \delta - b\lambda)\left(\frac{\alpha\delta\Gamma(a)\beta^b\Gamma(b)}{\Gamma(a+b)}\right)^\lambda \quad (21)$$

$$((-b\lambda + \delta\lambda + \delta))$$
$$\times {}_2F_1(-((a+b)\lambda), \lambda\alpha + \alpha - a\lambda; \lambda\alpha + \alpha + \delta - a\lambda - b\lambda + \delta\lambda + 1; 1 - \beta)$$
$$+ (\alpha\lambda - a\lambda + \alpha) {}_2F_1(-((a+b)\lambda), \lambda\alpha + \alpha - a\lambda + 1; \quad (22)$$
$$\lambda\alpha + \alpha + \delta - a\lambda - b\lambda + \delta\lambda + 1; 1 - \beta))\Bigg]$$
$$- \frac{\Gamma(\alpha)\Gamma(\delta)}{\Gamma(\alpha+\delta)}\Bigg], \quad (23)$$

on using Mathematica, and after some algebraic simplification.

Observe that for small distance between two probability distributions, it is expected that the minimum sample size required to discriminate them will be large. If not, then a small or moderate sample size is sufficient to discriminate between the probability distributions. It is assumed that the user/practitioner will specify in advance the PCS and the tolerance level in terms of the distance between the KW and the LNGB distributions. In case a tolerance level is specified (by means of some distance), the two distribution functions are not considered to be significantly different, if their distance does not exceed the tolerance level. Based on a given value of the PCS and a given tolerance level, one can determine the minimum sample size required to discriminate between the two distributions. Next, we are interested in finding the required sample size $n$ such that the PCS achieves a certain protection level $p$ for a stated tolerance level $D_1$. We explain the procedure under the null hypothesis that the true distribution is a two-parameter KW distribution. A similar procedure can mimicked for the LNGB distribution, and that is why it is omitted for brevity. To determine the sample size needed to attain at least a tolerance level $p$, we set $PCS_{KW}(\alpha, \delta) = p$. Hence, using the asymptotic result as obtained in Eq. (18), we get

$$\Phi\left(-\frac{\sqrt{n}M_{KW}(\alpha,\delta)}{\sqrt{Var_{KW}(\alpha,\delta)}}\right) = p.$$

Then, on solving for $n$, we have

$$n = \left[\frac{z_p^2 \times Var_{KW}(\alpha,\delta)}{M_{KW}(\alpha,\delta)}\right], \quad (24)$$

where $z_p$ is the 100$p$ th percentile point of the standard normal distribution.

Similarly, under the null hypothesis $H_{LNGB}$, and using the result in Eq. (17), the minimum sample size requirement will lead us to the following:

$$n = \left[\frac{z_p^2 \times Var_{LNGB}(a,b,\beta)}{M_{LNGB}(a,b,\beta)}\right], \quad (25)$$

Values of Eq. (24), for some representative values $\alpha$, corresponding to $\delta = 2.5$, and $p = 0.25, 0.55, 0.75$, are provided in Table 3. In Table 4, lists some values of Eq. (25) for some representative values of $a$, and with $b = 1.5$, and $\beta = 2.5$. We will briefly discuss how to use the PCS and the tolerance level in a practical setting. Suppose that an experimenter is interested in discriminating between the LNGB and KW distributions where the null hypothesis is $H_{KW}$. Further, suppose that the tolerance level is based on the Power divergence statistics, and it is fixed at 0.0211. Therefore, from the Table 3, one needs to take the sample size $n \geq 864$, for $p = 0.75$ to discriminate between the two distributions. For a more accurate result, under the hypothesis $H_{KW}$ (or $H_{LNGB}$), a greater range of $\alpha$ (as well as $a$ and $b$) will be required.

## 5. Simulation study

Here, we begin our discussion by providing an outline to carry out the simulation study. We are interested in comparing the asymptotic PCS's under the null hypotheses LNGB and KW with respect to the simulated probabilities based on Monte Carlo simulations. We begin with the case where the null hypothesis is $H_{KW}$. A similar procedure holds when the null hypothesis is a 3-parameter LNGB, and therefore is omitted for brevity.

**Table 3.** Values of $n$ and the Hellinger and the Power divergence distances between $KW(\alpha, \delta)$ and $LNGB(a, b, \beta)$ distribution for $\delta = 2.5$, and for some values of $\alpha$.

| $\alpha \rightarrow$ | 1.5 | 2 | 2.5 | 3 | 3.5 | 4 |
|---|---|---|---|---|---|---|
| n (p = 0.25) | 17 | 67 | 345 | 156 | 77 | 58 |
| n (p = 0.55) | 60 | 324 | 1034 | 598 | 348 | 257 |
| n (p = 0.75) | 167 | 864 | 2433 | 933 | 856 | 642 |
| $\mathbb{H}$ | 0.0022 | 0.0004 | 0.0001 | 0.0002 | 0.0004 | 0.0005 |
| PWD | 0.0364 | 0.0273 | 0.0211 | 0.0157 | 0.0123 | 0.0098 |

**Table 4.** Values of $n$ and the Hellinger and the Power divergence distances between $KW(\alpha, \delta)$ and $LNGB(a, b, \beta)$ distribution for $b = 1.5$, and $\beta = 2.5$ and for some values of $a$.

| $a \rightarrow$ | 0.25 | 0.35 | 0.45 | 1.25 | 1.45 | 2 |
|---|---|---|---|---|---|---|
| n (p = 0.25) | 14 | 59 | 425 | 159 | 63 | 48 |
| n (p = 0.55) | 64 | 338 | 1545 | 635 | 289 | 242 |
| n (p = 0.75) | 153 | 734 | 2163 | 1569 | 933 | 718 |
| $\mathbb{H}$ | 0.0022 | 0.0004 | 0.0001 | 0.0002 | 0.0004 | 0.0005 |
| PWD | 0.0314 | 0.0213 | 0.0151 | 0.0029 | 0.0017 | 0.0003 |

**Table 5.** Asymptotic probability under the null hypothesis $H_{KW}$.

| $\alpha \downarrow n \rightarrow$ | 25 | 40 | 70 | 85 | 100 | 150 | 400 |
|---|---|---|---|---|---|---|---|
| 0.2 | 0.6669 | 0.7291 | 0.7725 | 0.8058 | 0.8326 | 0.9137 | 0.9845 |
| 0.5 | 0.5755 | 0.6062 | 0.6293 | 0.6484 | 0.6649 | 0.7265 | 0.8296 |
| 0.9 | 0.5071 | 0.5100 | 0.5122 | 0.5141 | 0.5158 | 0.5223 | 0.5352 |
| 1.5 | 0.5365 | 0.5516 | 0.5631 | 0.5727 | 0.5812 | 0.6140 | 0.6766 |
| 2.0 | 0.5574 | 0.5809 | 0.5988 | 0.6137 | 0.6266 | 0.6761 | 0.7649 |
| 3.0 | 0.5717 | 0.6009 | 0.6229 | 0.6411 | 0.6570 | 0.7162 | 0.8270 |
| 5.0 | 0.5940 | 0.6254 | 0.6475 | 0.6650 | 0.6850 | 0.7500 | 0.8520 |

**Table 6.** Empirical probability under the null hypothesis $H_{KW}$.

| $\alpha \downarrow n \rightarrow$ | 25 | 40 | 70 | 85 | 100 | 150 | 400 |
|---|---|---|---|---|---|---|---|
| 0.2 | 0.7040 | 0.7370 | 0.7890 | 0.8120 | 0.8350 | 0.9280 | 0.9840 |
| 0.5 | 0.5760 | 0.6090 | 0.6400 | 0.6480 | 0.6640 | 0.7200 | 0.8270 |
| 0.9 | 0.4934 | 0.5002 | 0.4980 | 0.5072 | 0.5040 | 0.5018 | 0.5260 |
| 1.5 | 0.5380 | 0.5400 | 0.5500 | 0.5750 | 0.5730 | 0.6280 | 0.6790 |
| 2.0 | 0.5900 | 0.5830 | 0.5680 | 0.5990 | 0.6090 | 0.6930 | 0.7690 |
| 3.0 | 0.5828 | 0.6112 | 0.6256 | 0.6438 | 0.6562 | 0.7126 | 0.8146 |
| 5.0 | 0.5870 | 0.6221 | 0.6683 | 0.6799 | 0.6885 | 0.7665 | 0.8642 |

**Table 7.** Asymptotic probability under the null hypothesis $H_{LNGB}$.

| $\alpha \downarrow n \rightarrow$ | 25 | 40 | 70 | 85 | 100 | 150 | 400 |
|---|---|---|---|---|---|---|---|
| 0.2 | 0.7778 | 0.8602 | 0.9073 | 0.9369 | 0.9563 | 0.9922 | 0.9999 |
| 0.5 | 0.6458 | 0.6645 | 0.6788 | 0.6908 | 0.7013 | 0.7418 | 0.8171 |
| 0.9 | 0.5053 | 0.5074 | 0.5091 | 0.5105 | 0.5118 | 0.5166 | 0.5263 |
| 1.5 | 0.5266 | 0.5976 | 0.6161 | 0.6632 | 0.7194 | 0.7536 | 0.7908 |
| 2.0 | 0.6059 | 0.6383 | 0.6802 | 0.7518 | 0.7931 | 0.8186 | 0.8594 |
| 3.0 | 0.5944 | 0.6162 | 0.6877 | 0.7788 | 0.8599 | 0.9039 | 0.9520 |
| 5.0 | 0.6295 | 0.6417 | 0.6511 | 0.6589 | 0.6658 | 0.6926 | 0.7445 |

**Table 8.** Empirical probability under the null hypothesis $H_{LNGB}$.

| $\alpha \downarrow n \rightarrow$ | 25 | 40 | 70 | 85 | 100 | 150 | 400 |
|---|---|---|---|---|---|---|---|
| 0.2 | 0.8250 | 0.8400 | 0.8970 | 0.9220 | 0.9520 | 0.9930 | 0.9990 |
| 0.5 | 0.6360 | 0.6548 | 0.6654 | 0.6894 | 0.7038 | 0.7406 | 0.8116 |
| 0.9 | 0.5048 | 0.5246 | 0.5050 | 0.5190 | 0.5254 | 0.5264 | 0.5332 |
| 1.5 | 0.4624 | 0.5866 | 0.6104 | 0.6682 | 0.7088 | 0.7590 | 0.7824 |
| 2.0 | 0.6060 | 0.6240 | 0.6870 | 0.7280 | 0.7490 | 0.8130 | 0.8760 |
| 3.0 | 0.5950 | 0.6380 | 0.6700 | 0.7700 | 0.8600 | 0.8900 | 0.9330 |
| 5.0 | 0.5592 | 0.5880 | 0.6120 | 0.6204 | 0.6224 | 0.6658 | 0.7272 |

Let $M$ be the number of loops of the Monte Carlo simulation and $J = (J_1, \ldots, J_M)^T$ be a vector of length $M$. The steps, for each loop $\ell$, can be considered as follows:

1. Generate a random sample of size $n$ from the $KW(\alpha, \delta)$ distribution.
2. Find the MLEs of $(\alpha, \delta)$ and $(a, b, \beta)$ based on the KW and LNGB distributions, respectively.
3. Compute the observed value of the test statistic $W_n = \ell_{KW}(\widehat{\alpha}, \widehat{\delta}) - \ell_{LNGB}(\widehat{a}, \widehat{b}, \widehat{\beta})$.
4. If $W_n > 0$ take $J_\ell = 1$, otherwise $J_\ell = 0$.

Then, at the completion of the Monte Carlo simulation, the simulated PCS will be $\frac{\sum_{\ell=1}^{M} J_\ell}{M}$.

We also compute the PCS based on the asymptotic results derived in Section 2 and for the computations, statistical software R is utilized (R Core Team 2019).

In Tables 5–6, a we present the asymptotic and simulated PCS values under the null hypothesis that the true distribution of the data is a two-parameter KW for $\alpha = 0.2, 0.5, 0.9, 1.5, 2.0, 3.0, 5.0$, and $n = 25, 40, 70, 85, 100, 150, 400$ with a fixed $\delta = 2.5$.

From the values obtained in Tables 5–6, it is quite clear that there is a good agreement between the asymptotic and empirical probabilities, mainly for moderate and large sample sizes.

We also observe that, when $\alpha$ approaches 1, the PCSs approaches 0.5. This was expected as $\alpha$ goes to 0, both KW and LNGB distributions converge to the same law. Another expected result we observed is that when $n$ increases the PCS approaches one.

In Tables 7–8, we present the asymptotic and simulated PCS values under the null hypothesis that the true distribution of the data is a three-parameter LNGB for $a = 0.2, 0.5, 0.9, 1.5, 2.0, 3.0, 5.0$, and $n = 25, 40, 70, 85, 100, 150, 400$ for a fixed $b = 1.25$, $\beta = 1.5$. In this case we also observe a good agreement between the asymptotic and empirical PCS values. When $a$ is close to one, the PCS values are close to 0.5, and as $n$ increases, the probabilities goes to one, as expected and discussed in the previous case.

## 6. Real data application

### 6.1. Application 1: HIV data set

Here, we consider a Brazilian HIV data set to illustrate the feasibility of the proposed methodology described in Section 3. developed in this article. This data set was originally used by Louzada et al. (2012) and was re-analyzed in Arnold and Ghosh (2017) in the context of copula based construction of bivariate KW models. This data set contains information on patients (older than 18 years of age) two different stages of admission with HIV seen at the Servicio de Doencas Infecciosas e Parasitarias (DIP), Universidade Federal do Triangulo Mineiro (UFTM), Brazil, diagnosed with HIV between January 1996 and December 1999. We consider the following variable: $Y_1$ : The proportion of timely admitted patients (first admission, preferably at the first stage of HIV, i.e., chance of partial/and or complete cure to HIV) for each month during the time interval January 1996–December 1998. Overall, we have 48 of such observations corresponding to $Y_1$.

We conjecture at this point that both the two parameter KW distribution and the three parameter LNGB distribution are natural candidates to fit this data set. On applying the proposed methodology, we report the following: The MLEs of the parameters for the KW and LNGB distributions are $(\widehat{\alpha}, \widehat{\delta}) = (1.3458, 2.6433)$ and $(\widehat{a}, \widehat{b}, \widehat{\beta}) = (0.6783, 0.5342, 1.2923)$, respectively. The test statistic equals $W_n = 132.4324 - 128.6536 = 3.7788 > 0$, indicating that the LNGB model should be selected according to the AIC (Akaike Information Criterion). Under the hypothesis that the data come from a $KW(1.3458, 2.6433)$ distribution, we obtain the following estimated quantities: $M_{KW}(\widehat{\alpha}, \widehat{\delta}) = 0.3211$, and $\text{Var}_{KW}(\widehat{\alpha}, \widehat{\delta}) = 0.4329$. Therefore, from Eq. (18), we have $\text{PCS}_{KW}(\widehat{\alpha}, \widehat{\delta}) = 0.0003608$. Similarly, under the null hypothesis that the data come from a LNGB distribution, we have $M_{LNGB}(\widehat{a}, \widehat{b}, \widehat{\beta}) = 0.0127$, and $\text{Var}_{LNGB}(\widehat{a}, \widehat{b}, \widehat{\beta}) =$ 0.4522. Therefore, from Eq. (17), $\text{PCS}_{LNGB}\left(\widehat{a},\widehat{b},\widehat{\beta}\right) = 0.5520$. Therefore, the probability of correct selection (based on the asymptotic results) is at least equal to $\min\{0.0003608, 0.5520\} = 0.0003608$. Since the PCS is maximum under the hypothesis $H_{LNGB}$, we select the LNGB distribution. Based on the simulated PCS values, we obtain the same conclusion.

### 6.2. Application 2: On modeling arthritic pain relief times data

We consider a data set from a medical field that has been analyzed previously. The data set reports results from a clinical trial that was performed to assess the efficacy of an analgesic. These data show relaxation periods (in hours) of 50 arthritic patients taking a fixed dosage of certain drug, for details on this data set, see Wingo (1983). On applying the proposed methodology, we report the following: The MLEs of the parameters for the KW and LNGB distributions are (for this data set) $\left(\widehat{\alpha},\widehat{\delta}\right) = (2.465, 4.324)$, and $\left(\widehat{a},\widehat{b},\widehat{\beta}\right) = (0.9534, 1.2438, 1.4562)$, respectively. The test statistic equals $W_n = 156.3893 - 137.4528 = 18.9365 > 0$, indicating that the LNGB model should be selected according to the AIC (Akaike Information Criterion). Under the hypothesis that the data come from a $KW(2.465, 4.324)$ distribution, we obtain the following estimated quantities: $M_{KW}\left(\widehat{\alpha},\widehat{\delta}\right) = 0.5237$, and $\text{Var}_{KW}\left(\widehat{\alpha},\widehat{\delta}\right) = 0.4982$. Therefore, from Eq. (18), we have $\text{PCS}_{KW}\left(\widehat{\alpha},\widehat{\delta}\right) = 0.00027936$. Similarly, under the null hypothesis that the data come from a LNGB distribution, we obtain $M_{LNGB}\left(\widehat{a},\widehat{b},\widehat{\beta}\right) = 0.03451$, and $\text{Var}_{LNGB}\left(\widehat{a},\widehat{b},\widehat{\beta}\right) = 0.3817$. Therefore, from Eq. (17), the associated PCS will be $\text{PCS}_{LNGB}\left(\widehat{a},\widehat{b},\widehat{\beta}\right) = 0.6535$. Therefore, the probability of correct selection (based on the asymptotic results) is at least equal to $\min\{0.00027936, 0.6535\} = 0.00027936$. Since the PCS is maximum under the hypothesis $H_{LNGB}$, we select the LNGB distribution. Based on the simulated PCS values, we arrive at the same conclusion.

### 7. Concluding remarks

Discriminating between two absolutely continuous probability distributions in not new in the literature. However, not much work has been done to study on the discrimination between two probability distributions for modeling continuous bounded data. In this article we discuss and explore the strategy of discriminating between a three parameter generalized beta distribution proposed by Libby-Novick and a two parameter Kumaraswamy distribution. It can be conjectured at this point that similar strategy can be adopted in higher dimension, such as between a multivariate Kumaraswamy distribution and a multivariate generalized beta distribution. The only hindrance it could be in pursuing this research is that the practitioner needs to find a real-life motivation.

### Disclosure statement

The author do not have any conflict of interest in preparing this manuscript.

### Data availability statement

The complete data references are adequately mentioned in the paper.

### ORCID

Indranil Ghosh https://orcid.org/0000-0001-7910-8919

### Appendix A

Here, we provide the mathematical details on the derivation of the mean and variance of the statistic $W_n$ under the null hypothesis that LNGB is the true distribution of the data.

Let us define

$$\Delta_{LNGB}(\alpha, \delta)$$
$$= E_{LNGB}\left[\log f_{KW}(X; \alpha, \delta)\right]$$
$$= E_{LNGB}\left[\log \alpha + \log \delta + (\alpha - 1)\log X + (\delta - 1)\log\left[1 - X^\alpha\right]\right]$$
$$= \log \alpha + \log \delta + (\alpha - 1)\frac{\beta^a}{B(a,b)}\int_0^1 x^{a-1}(1-x)^{b-1}$$
$$\log x\left[1 - (1-\beta)x\right]^{a+b} dx$$
$$+ (\delta - 1)\frac{\beta^a}{B(a,b)}\int_0^1 x^{a-1}(1-x)^{b-1}\log\left[1 - x^\alpha\right]$$
$$\left[1 - (1-\beta)x\right]^{a+b} dx$$
$$= \log \alpha + \log \delta + (\alpha - 1)\frac{\beta^a}{B(a,b)}$$
$$\left[\sum_{k=0}^{\infty}(-1)^k(1-\beta)^k B(a+b+k)\left(\Psi(a+k) - \Psi(a+b+k)\right)\right]$$
$$- (\delta - 1)\frac{\beta^a}{B(a,b)}\sum_{k=0}^{\infty}\sum_{\ell=0}^{\infty}\frac{(-1)^k}{\ell}\binom{a+b}{k}(1-\beta)^k$$
$$\times \left\{(\beta - 1)\Gamma(b)\Gamma(a+b+k)\,_3F_2\big((1,1,1+a+k),\right.$$
$$\left.(2,1+a+b+k), 1-\beta\big)\right\}, \tag{A1}$$

on using Mathematica, and after some algebraic simplification. Consequently, we have $\tilde{\alpha}, \tilde{\delta}$ which are obtained as the solutions of the following system of non-linear equations

$$\left(\frac{\partial \Delta_{LNGB}(\alpha, \delta)}{\partial \alpha}, \frac{\partial \Delta_{LNGB}(\alpha, \delta)}{\partial \delta}\right)^T = (0,0)^T, \tag{A2}$$

equivalently

$$\frac{1}{\tilde{\alpha}} + \frac{\beta^a}{B(a,b)}\left[\sum_{k=0}^{\infty}(-1)^k(1-\beta)^k B(a+b+k)\right.$$
$$\left.(\Psi(a+k) - \Psi(a+b+k))\right] = 0, \tag{A3}$$

$$\frac{1}{\tilde{\delta}} - \frac{\beta^a}{B(a,b)}\sum_{k=0}^{\infty}\sum_{\ell=0}^{\infty}\frac{(-1)^k}{\ell}\binom{a+b}{k}(1-\beta)^k$$
$$\times \left\{(\beta - 1)\Gamma(b)\Gamma(a+b+k)\,_3F_2\big((1,1,1+a+k),\right.$$
$$\left.(2,1+a+b+k), 1-\beta\big)\right\} = 0. \tag{A4}$$

In our case, we have

$$M_{LNGB}$$
$$= E_{LNGB}\left[\log f_{LNGB}(X; a,b,\beta) - \log f_{KW}(X; \tilde{\alpha}, \tilde{\delta})\right]$$
$$= \int_0^1 \log f_{LNGB}(X; a,b,\beta) f_{LNGB}(X; a,b,\beta) dx$$
$$- \int_0^1 \log f_{KW}(X; \tilde{\alpha}, \tilde{\delta}) f_{LNGB}(X; a,b,\beta) dx$$
$$= a\log\beta - \log B(a,b) + (a-1)\sum_{k=0}^{a+b}(-1)^k\binom{a+b}{k}(1-\beta)^k$$
$$\{(a-1)B(b, a+k)(\Psi(a+k) - \Psi(a+b+k))$$
$$+ (b-1)B(b, a+k)(\Psi(b) - \Psi(a+b+k))$$
$$+ (a+b)(\beta-1)\Gamma(b)\Gamma(1+a+k)\,_3F_2\big((1,1,1+a+k),$$
$$(2, 1+a+b+k), 1-\beta\big)\}$$
$$- \log \tilde{\alpha} - \log \tilde{\delta} - (\tilde{\alpha} - 1)\sum_{k=0}^{a+b}(-1)^k\binom{a+b}{k}(1-\beta)^k$$
$$B(b, a+k)(\Psi(b) - \Psi(a+b+k))$$
$$- (\tilde{\delta} - 1)\sum_{\ell=0}^{\tilde{\delta}-1}\sum_{k=0}^{a+b}(-1)^{k+\ell}\binom{a+b}{k}\binom{\tilde{\delta}-1}{\ell}$$
$$(1-\beta)^k B(a+k+\alpha\ell, b). \tag{A5}$$

Similarly,

$$Var_{LNGB}$$
$$= Var_{LNGB}\left[\log f_{LNGB}(X; a,b,\beta) - \log f_{KW}(X; \tilde{\alpha}, \tilde{\delta})\right]$$
$$= Var_{LNGB}\left[\left(a\log\beta - \log B(a,b) + (a-1)\log X\right.\right.$$
$$\left.+ (b-1)\log(1-X) + (a+b)\log(1-(1-\beta)X)\right) -$$
$$\left.\left(\log\left(\tilde{\alpha}\tilde{\delta}\right) + (\tilde{\alpha} - 1)\log X + (\tilde{\delta} - 1)\log\left[1 - X^{\tilde{\alpha}}\right]\right)\right]$$
$$= (a - \tilde{\alpha})^2 Var_{LNGB}(\log X) + (b-1)^2 Var_{LNGB}(\log(1-X))$$
$$+ (a+b)^2 Var_{LNGB}\left[\log(1-(1-\beta)X)\right]$$
$$+ (b-1)^2 Var_{LNGB}\left[\log(1-X)\right] + (\tilde{\delta} - 1)^2 Var_{LNGB}\left[\log\left[1 - X^{\tilde{\alpha}}\right]\right]$$
$$- 2(a-1)(\tilde{\delta} - 1) Cov_{LNGB}\left[\log X, \log(1-X)\right]$$
$$- 2(a+b)(\tilde{\delta} - 1) Cov_{LNGB}\left[\log(1-(1-\beta)X), \log(1-X^{\tilde{\alpha}})\right]$$
$$- 2(a - \tilde{\alpha})(b-1) Cov_{LNGB}\left[\log X, \log(1-X)\right]. \tag{A6}$$

## Appendix B

In this section, we provide the details of the mathematical derivation of certain quantities utilized in this manuscript. We begin with the following quantity

1.
$$\Delta_{KW}(a,b,\beta)$$
$$= E_{KW}\left[\log f_{LNGB}(X; a,b,\beta)\right]$$
$$= a\log\beta - \log B(a,b) + \alpha\delta(a-1)I_1$$
$$+ \alpha\delta(b-1)I_2 - \alpha\delta(a+b)I_3, \quad \text{say,} \tag{B1}$$

Next, on using Mathematica, we consider evaluation each of the three integrals separately, which are as follows:

$$I_1 = \int_0^1 \left[\log x\right] x^{\alpha - 1}\left(1 - x^\alpha\right)^{\delta - 1} dx$$
$$= \sum_{j=0}^{\delta-1}(-1)^j\binom{\delta-1}{j}\int_0^1 \log(x) x^{\alpha(1+j)-1} dx$$
$$= \sum_{j=0}^{\delta-1}(-1)^j\binom{\delta-1}{j}\frac{1}{\alpha^2(j+1)^2}, \tag{B2}$$

if $\delta$ is not an integer, the sum will go up to infinity.

Next,

$$I_2 = \int_0^1 \log(1-x) x^{\alpha - 1}\left(1 - x^\alpha\right)^{\delta - 1} dx$$

$$= \sum_{j=0}^{\delta-1}(-1)^j\binom{\delta-1}{j}\int_0^1 \log(1-x)\,x^{\alpha(1+j)-1}dx$$

$$= -\sum_{j=0}^{\delta-1}(-1)^j\binom{\delta-1}{j}\frac{H_{j\alpha+\alpha}}{\alpha+\alpha j}, \quad (B3)$$

where $H_m$ is the harmonic number and if $\delta$ is not an integer, the sum will go up to infinity.

Next,

$$I_3 = \int_0^1 [\log(1-(1-\beta)x)]\,x^{\alpha-1}(1-x^\alpha)^{\delta-1}dx$$

$$= \sum_{j=0}^{\delta-1}(-1)^j\binom{\delta-1}{j}\log(1-(1-\beta)x)\,x^{\alpha(1+j)-1}dx$$

$$= \sum_{j=0}^{\delta-1}(-1)^j\binom{\delta-1}{j}\frac{\log(\beta)+(1-\beta)^{\alpha(-j-1)}B_{1-\beta}(j\alpha+\alpha+1,0)}{\alpha(j+1)}, \quad (B4)$$

if $\delta$ is not an integer, the sum will go up to infinity. On substitution Eqs. (B2)–(B4), in Eq. (B1), one may obtain a closed form expression of the expectation.

2. Next, we want to obtain a closed form expression of the following quantity

$$Var_{KW}(\alpha, \delta)$$
$$= Var_{KW}\left[\log f_{LNGB}(X; \tilde{a}, \tilde{b}, \tilde{\beta}) - \log f_{KW}(X; \alpha, \delta)\right]$$
$$= (\tilde{a}+\alpha-1)^2 Var^{(1)}_{KW}[\log X]$$
$$+ (\tilde{b}-1)^2 Var^{(1)}_{KW}[\log(1-(1-\beta)X)]$$
$$+ (\tilde{\beta}-1)^2 Var^{(1)}_{KW}[\log(1-X^\alpha)]$$
$$+ 2(\tilde{a}+\alpha-1)(\tilde{b}-1) Cov_{KW}[\log X, \log(1-(1-\beta)X)]$$
$$+ 2(\tilde{a}+\alpha-1)(\tilde{\beta}-1) Cov_{KW}[\log X, \log(1-X^\alpha)]$$
$$+ 2(\tilde{\beta}-1)(\tilde{b}-1) Cov_{KW}\left[\log(1-X^\alpha), \log(1-(1-\tilde{\beta})X)\right]. \quad (B5)$$

One can figure out the fact that while computing the covariance, some of the individual expectation terms have already been obtained earlier. For example, consider

$$Cov_{KW}\left(\log X, \log(1-(1-\beta)X)\right)$$
$$= E_{KW}[(\log X)(\log(1-(1-\beta)X))]$$
$$- \{E_{KW}[\log X]\}\{E_{KW}(\log(1-(1-\beta)X))\}. \quad (B6)$$

Next, consider the following:

$$E_{KW}\left((\log X)(\log(1-(1-\beta)X))\right)$$
$$= \sum_{j=0}^{\delta-1}(-1)^j\binom{\delta-1}{j}\int_0^1 \log(x)\log(1-(1-\beta)x)\,x^{\alpha-1}$$
$$(1-x^\alpha)^{\delta-1}dx$$
$$= \left[\alpha(\beta-1)(j+1)(\Phi(1-\beta,2,j\alpha+\alpha+1)\right.$$

$$\left.+\Gamma(j\alpha+\alpha)\,{}_2\tilde{F}_1(1,j\alpha+\alpha+1;j\alpha+\alpha+2;1-\beta))-\log(\beta)\right]$$
$$\times \left(\alpha^2(j+1)^2\right)^{-1}, \quad (B7)$$

if $\delta$ is not an integer, the sum will go up to infinity. Therefore, substituting Eqs. (B6), (B7) in (B5), one can get the closed form expression of the associated covariance term.

Next, consider

$$Cov_{KW}\left[\log X, \log(1-X^\alpha)\right]$$
$$= E_{KW}[(\log X)(\log(1-X^\alpha))]$$
$$- \{E_{KW}[\log X]\}\{E_{KW}(\log(1-X^\alpha))\}. \quad (B8)$$

Now, consider,

$$E_{KW}\left[(\log X)(\log(1-X^\alpha))\right]$$
$$= \sum_{j=0}^{\delta-1}(-1)^j\binom{\delta-1}{j}\int_0^1 \log(x)\log(1-x^\alpha)\,x^{\alpha-1}(1-x^\alpha)^{\delta-1}dx$$
$$= \sum_{j=0}^{\delta-1}(-1)^j\binom{\delta-1}{j}$$
$$\left(\frac{\gamma j+(j+1)^2(-\psi^{(1)}(j+2))+(j+1)\psi^{(0)}(j+1)+\gamma+1}{\alpha^2(j+1)^3}\right), \quad (B9)$$

where $\gamma$ is the Euler's gamma function and $\psi^{(m)}(t)$ is defined earlier, and using Mathematica. Therefore, substituting Eqs. (B8), (B9) in (B6), one can get the closed form expression of the associated covariance term.

Let us define

$$\Delta_{KW}(a,b,\beta)$$
$$= E_{KW}\left[\log f_{LNGB}(X; a, b, \beta)\right]$$
$$= a\log\beta - \log B(a,b) + \alpha\delta(a-1)\int_0^1 \log x\,x^{\alpha-1}(1-x^\alpha)^{\delta-1}dx$$
$$+ \alpha\delta(b-1)\int_0^1 \log(1-x)\,x^{\alpha-1}(1-x^\alpha)^{\delta-1}dx$$
$$- \alpha\delta(a+b)\int_0^1 [\log(1-(1-\beta)x)]\,x^{\alpha-1}(1-x^\alpha)^{\delta-1}dx$$
$$= a\log\beta - \log B(a,b) - \alpha\delta(a-1)\sum_{j=0}^{\delta-1}(-1)^j\binom{\delta-1}{j}\frac{1}{\alpha^2(j+1)^2}$$
$$- \alpha\delta(b-1)\sum_{j=0}^{\delta-1}(-1)^j\binom{\delta-1}{j}\frac{H_{j\alpha+\alpha}}{\alpha+\alpha j}$$
$$- \alpha\delta(a+b)\sum_{j=0}^{\delta-1}(-1)^j\binom{\delta-1}{j}$$
$$\frac{\log(\beta)-(1-\beta)^{\alpha(j+1)}B_{1-\beta}(j\alpha+\alpha+1,0)}{\alpha(j+1)}, \quad (B10)$$

where $H_m$ is the harmonic number and if $\delta$ is not an integer, the sum will go up to infinity, and the expression is obtained on using Mathematica.

Therefore, we have, $\tilde{a}, \tilde{b}, \tilde{\beta}$ which are obtained as the solutions of the system of non-linear equations.

$$\left(\frac{\partial \Delta_{KW}(a,b,\beta)}{\partial a}, \frac{\partial \Delta_{KW}(a,b,\beta)}{\partial b}, \frac{\partial \Delta_{KW}(a,b,\beta)}{\partial \beta}\right)^T = (0,0,0)^T. \quad (B11)$$

Due to the complex nature of the associated likelihood equations, we do not provide them here. It is available upon request to the author.

Next, under $H_{KW}$, we compute the mean and variance of the random variable $W_n = \log f_{LNGB}(X; \tilde{a}, \tilde{b}, \tilde{\beta}) - \log f_{KW}(X; \alpha, \delta)$ which will be denoted by $M_{KW}(\alpha, \delta)$ and $Var_{KW}(\alpha, \delta)$, respectively.

In our case

$M_{KW}(\alpha, \delta)$

$= E_{KW}\left[\log f_{LNGB}\left(X; \tilde{a}, \tilde{b}, \tilde{\beta}\right) - \log f_{KW}(X; \alpha, \delta)\right]$

$= \tilde{a} \log \tilde{\beta} - \log B\left(\tilde{a}, \tilde{b}\right) - \log \alpha - \log \delta + (\tilde{a} + \alpha - 1) E_{KW}[\log X]$

$+ \left(\tilde{b} - 1\right) E_{KW}\left[\log\left(1 - (1 - \tilde{\beta})\right)X\right] + \left(\tilde{\beta} - 1\right) E_{KW}\left[\log\left(1 - X^\alpha\right)\right]$

$= \tilde{a} \log \tilde{\beta} - \log B\left(\tilde{a}, \tilde{b}\right) - \log \alpha - \log \delta + (\tilde{a} + \alpha - 1) E_{KW}[\log X]$

$- (\tilde{a} + \alpha - 1) \sum_{j=0}^{\delta-1} (-1)^j \binom{\delta - 1}{j} \frac{1}{\alpha^2(j+1)^2}$

$+ \left(\tilde{b} - 1\right) \sum_{j=0}^{\delta-1} (-1)^j \binom{\delta - 1}{j}$

$\frac{\log(\tilde{\beta}) + (1 - \tilde{\beta})^{\alpha(-j-1)} B_{1-\tilde{\beta}}(j\alpha + \alpha + 1, 0)}{\alpha(j+1)}$

$+ \left(\tilde{\beta} - 1\right) \sum_{j=0}^{\delta-1} (-1)^j \binom{\delta - 1}{j} \frac{H_{j+1}}{\alpha(1+j)}, \quad \text{(B12)}$

where $H_m$ is the haromonic number and if $\delta$ is not an integer, the sum will go up to infinity, and using `Mathematica` and after some algebraic simplification.

Again,

$Var_{KW}(\alpha, \delta)$

$= Var_{KW}\left[\log f_{LNGB}(X; \tilde{a}, \tilde{b}, \tilde{\beta}) - \log f_{KW}(X; \alpha, \delta)\right]$

$= (\tilde{a} + \alpha - 1)^2 Var_{KW}[\log X] + \left(\tilde{b} - 1\right)^2$

$\quad Var_{KW}[\log(1 - (1 - \beta)X)]$

$+ \left(\tilde{\beta} - 1\right)^2 Var_{KW}\left[\log\left(1 - X^\alpha\right)\right]$

$+ 2(\tilde{a} + \alpha - 1)\left(\tilde{b} - 1\right) Cov_{KW}\left[\log X, \log(1 - (1 - \beta)X)\right]$

$+ 2(\tilde{a} + \alpha - 1)\left(\tilde{\beta} - 1\right) Cov_{KW}\left[\log X, \log\left(1 - X^\alpha\right)\right]$

$+ 2\left(\tilde{\beta} - 1\right)\left(\tilde{b} - 1\right) Cov_{KW}\left[\log\left(1 - X^\alpha\right), \log\left(1 - (1 - \tilde{\beta})X\right)\right].$

(B13)